# A Study of Multifractal Spectra and Renyi Dimensions in 14.5A GeV/c $^{28}$Si-Nucleus Collisions

N. Ahmad[1*], A. Kamal[2], M. M. Khan[3], Hushnud[1], A. Tufail[4]

[1]Department of Physics, Aligarh Muslim University, Aligarh, India
[2]Department of Basic Science, Deanship of Prepratory Year, Umme Al-Qura University, Makkah, KSA
[3]Department of Applied Physics, Aligarh Muslim University, Aligarh, India
[4]Applied Science Section, University Polytechnic, Aligarh Muslim University, Aligarh, India
Email: [*]nazeerahmadna@gmail.com

## Abstract

A systematic analysis of the data on 14.5A GeV/c $^{28}$Si-nucleus collisions is carried out to investigate the behaviours of Renyi dimensions, $D_q$ and Multifractal Spectral Function, $f(\alpha_q)$. The Renyi dimensions, $D_q$ are observed to decrease with increasing order of the moments, $q$. However, the Multifractal Spectra are concave downwards with their maxima occurring around $\alpha_q = 1.21 \pm 0.01$. A continuous curve representing Multifractal Spectral Function, $f(\alpha_q)$, characterizes manifestation of fluctuations in the rapidity space.

## 1. Introduction

Fluctuations of dynamical nature were observed for the first time in the JACEE event [1] having unusually very high multiplicity. The study of non-statistical fluctuations has since then attracted a great deal of attention of high energy physicists due to possibility of disentangling some important information about the mechanism of multiparticle production in relativistic hadronic and nuclear collisions. It has been observed [2] [3] that large particle density fluctuations in small rapidity bins exhibit self-similar behaviour. However, the origin of multifractality still remains largely unexplained. Self-similar behaviour of particles produced is closely linked to multifractality, which is a consequence of cascading mechanism of particle production in these collisions. Power-law behaviour of Scaled Factorial Moments indeed exhibits fractal patterns [4] [5] in the multiparticle dynamics of the final state of the collisions. Therefore, the issue of self-similarity under the perspective of fractal properties of the hadronic matter in multiparticle production needs to be adequately addressed. To investigate quantitatively any single particle density distribution in the framework of multifractal characteristics, a method has been developed [6]. This method has been successively applied in studying intermittent behaviour of particle density distributions over a wide range of pseudorapidity bin widths. In the present study, an attempt is made to investigate multifractal characteristics of the secondary particles produced in 14.5A GeV/c $^{28}$Si-nucleus collisions.

---

[*]Corresponding author.

## 2. Mathematical Formalism

In order to investigate various interesting features of multifractal moments in a given pseudorapidity range, $\Delta \eta$, $\Delta \eta (= \eta_{max} - \eta_{min})$, is divided into $M_0$ bins of width $\delta \eta = \Delta \eta / M_0$. As some of the bins may be empty, the number of non-empty bins is represented by $M$. According to Hwa [6], mutltifractal moments, $G_q$ are defined as:

$$G_q = \sum_{j=1}^{M} p_j^q \quad (1)$$

where $p_j = n_j/n$, such that $n = n_1 + n_2 + \cdots + n_M$ and $n_j$ denotes the number of particles $j^{th}$ bin. Here $q$ represents a real number, both positive and negative. The summation is performed over only non-empty bins and vertical average of horizontal moments is calculated from:

$$\langle Gq \rangle = (1/N) \sum_{1}^{N} G_q \quad (2)$$

where $N$ stands for the total number of events in a given data sample. If there occurs self-similarity in the production of particles, $Gq$ moments can be expressed in the form of a power-law as:

$$\langle Gq \rangle = (\delta \eta)^{T_q} \quad (3)$$

where $T_q$ are the mass exponents and may be determined from the observed linear dependence of $\ln \langle Gq \rangle$ on $-\ln \delta \eta$ using Equation (3). The multifractal spectrum function, $f(\alpha_q)$, and Renyi dimensions, $D_q$, are, therefore, used to examine fractal nature of particle emitting source in the analysis of the data. The Renyi dimensions are sometimes referred to as generalized dimensions. For nuclear collisions, $D_q$ are related to the slope parameter, $Tq$, as [5]:

$$Dq = (Tq)/(q-1) \quad (4)$$

However, $Tq$, $\alpha_q$ and $f(\alpha_q)$ may be related in the following function:

$$\alpha_q = (dTq)/dq \quad (5)$$

and

$$f(\alpha_q) = q\alpha_q - Tq \quad (6)$$

## 3. Experimental Details

In the present study a stack of ILFORD G5 emulsion exposed to 14.5A GeV/c $^{28}$Si nuclei from AGS, BNL has been used. A random sample of 555 interactions with $n_h \geq 0$, where $n_h$ represents the number of charged particles produced in an interactions with relative velocity, $\beta \leq 0.7$, is analyzed. In order to compare the experimental results with the corresponding values predicted by Lund Monte Carlo Model, FRITIOF, 5000 events with the same descriptions generated by FRITIOF, are also analyzed. For studying target mass dependence of various important parameters linked to multifractality, the given sample of collisions is divided in three categories: 1) of interactions due to CNO group; 2) interactions due to AgBr group and 3) entire sample.

## 4. Results and Discussion

The slope parameter, $T_q$, for the linear region of the $\ln\langle G_q \rangle$ versus $-\ln(\delta\eta)$ plot for each data set is calculated and plotted as a function of $q$ in **Figure 1**. It may be seen from the figure that $T_q$ increases significantly with increasing $q$ for $q < 0$, whereas it is observed to flatten with increasing $q$ for $q > 0$. The Renyi dimensions, $D_q$, and multfractal spectral function, $f(\alpha_q)$, are calculated using Equations (2)-(4) for both the experimental and FRITIOF data. Shown in **Figure 2** and **Figure 3** are $D_q$ versus $q$ and $f(\alpha_q)$ versus $\alpha_q$ plots for the experimental and FRITIOF data. It is clear from **Figure 2** that Renyi dimensions, $D_q$, decrease with increasing order of the moments, $q$, from −6 to 6, which indicates the presence of multifractality in multiparticle production in 14.5A GeV/c $^{28}$Si-nucleus collisions.

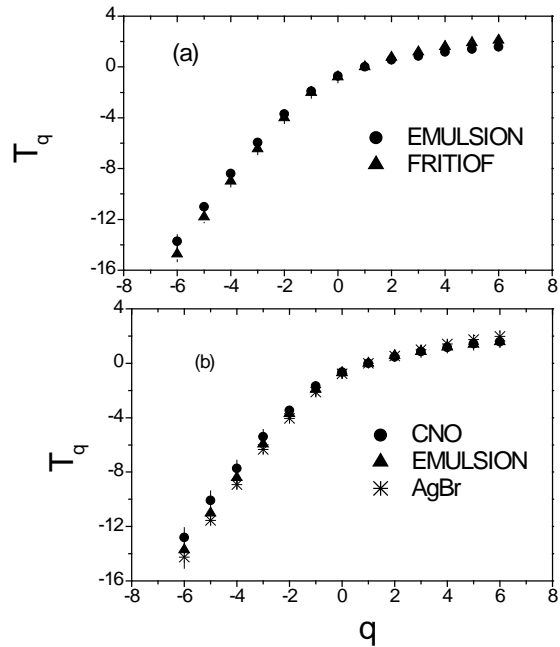

**Figure 1.** Variations of $T_q$ with $q$ in 14.5A GeV/c $^{28}$Si-nucleus collisions.

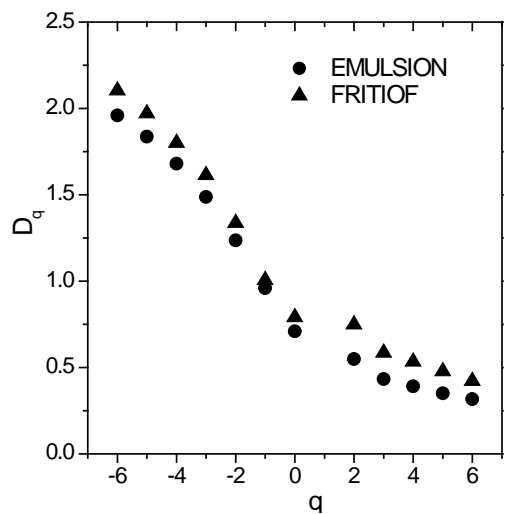

**Figure 2.** Variations of $D_q$ with $q$ in 14.5A GeV/c $^{28}$Si-nucleus collisions for the experimental and FRITIOF generated events.

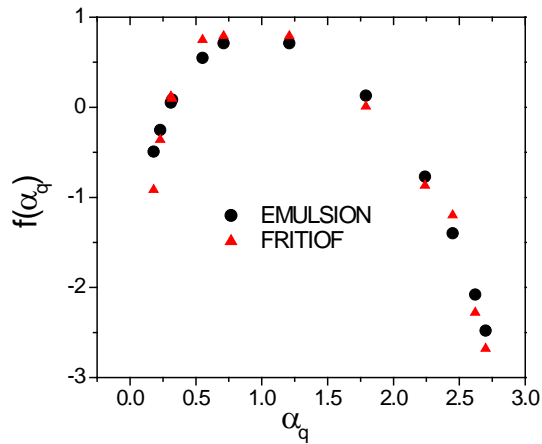

**Figure 3.** Variations of $f(\alpha_q)$ with $\alpha_q$ in 14.5A GeV/c $^{28}$Si-nucleus collisions for the experimental and FRITIOF generated events.

Variations of multifractal spectral function, $f(\alpha_q)$, with $\alpha_q$ for the FRITIOF and experimental data are displayed in **Figure 3**. It is seen from the figure that the value of Spectral Function, $f(\alpha_q)$, lies between −2.48 and 0.71 and −2.68 and 0.79 for the experimental and FRITIOF generated data respectively. The spectra are concave downwards centered around $\alpha_q = 1.21$ corresponding to $q = 0$.

In order to examine whether Renyi dimensions, $D_q$ and Multifractal Spectral Function $f(\alpha_q)$, depend on target mass, $D_q$ and $f(\alpha_q)$ are plotted as a function of $q$ and $\alpha_q$ respectively for the interactions of 14.5 A GeV/c $^{28}$Si nuclei with CNO, emulsion and AgBr targets and are shown in **Figure 4** and **Figure 5** respectively. Renyi dimensions are found to be relatively higher for the interactions due to heavier targets for each order of the moments. Although the effect appears to be rather more pronounced for $q < 0$. One of the possible reasons for Renyi dimensions being higher for the interactions due to heavier targets may be attributed to the fact that average multiplicity increases with increasing target size.

From $f(\alpha_q)$ versus $\alpha_q$ plot one may come to the following important conclusions:
1. In each case, Multifractal Spectral Function, $f(\alpha_q)$, is represented by a continuous curve, thus characterizing a quantitative manifestation of the fluctuations in the rapidity space.
2. The spectra are concave downwards in shape. However, for none of the two data sets, the spectrum has sharp peak, indicating thereby that pseudorapidity density distribution is not smooth.
3. The value of $f(\alpha_q)$ is relatively smaller for the targets of higher mass and the spectrum tends to become wider with increasing target size.

The $f(\alpha_q)$ spectrum may be used to estimate Renyi dimensions, $D_q$, which is the basic characteristics of any fractal structure. The values of $D_q$ are estimated [5]-[7] using:

$$D_0 = f(\alpha_q = 0), \quad D_1 = f(\alpha_q = 1), \quad \text{and} \quad D_2 = 2\alpha_{q=2} - f(\alpha_{q=2}) \tag{7}$$

The values of Renyi's dimensions, $D_0$, $D_1$ and $D_2$, which are regarded to be the most sensitive to the production mechanism, are calculated for the experimental and FRITIOF simulated data. The values of $D_0$, $D_1$ and $D_2$ are presented in **Table 1**. It may be seen from the table that values of $D_0$, $D_1$ and $D_2$ compare fairly well with the corresponding results obtained for the FRITIOF generated data.

## 5. Conclusion

From the analysis of 14.5A GeV/c $^{28}$Si-nucleus collisions performed in terms of multifractal moments, $G_q$, Multifractal Spectral Function, $f(\alpha_q)$, when plotted against $\alpha_q$, is observed to be concave in shape with a maximum occurring at around $f(\alpha_q) \sim 0.71 \pm 0.01$ at $\alpha_q = 1.21 \pm 0.01$. Furthermore, $f(\alpha_q)$ in the entire pseudorapidity space manifests self-similarity. The decreasing trend in the values of $D_q$ with increasing $q$ confirms the presence of multifractality in the collisions considered in the present study. Weak target mass dependence of $D_q$ and $f(\alpha_q)$ is revealed. Finally, Renyi dimensions, $D_0$, $D_1$ and $D_2$, obtained for the experimental data are found to be in excellent agreement with the corresponding values for the FRITIOF generated events. It may be remarked that, a similar study of nuclear interactions at still higher energies, including RHIC and LHC, may provide invaluable information regarding the mechanism of multiparticle production.

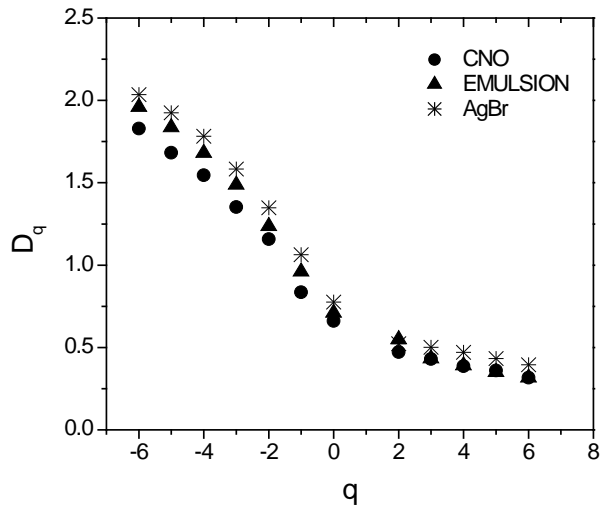

**Figure 4.** Variations of $D_q$ with $q$ in 14.5A GeV/c $^{28}$Si-nucleus collisions for various targets.

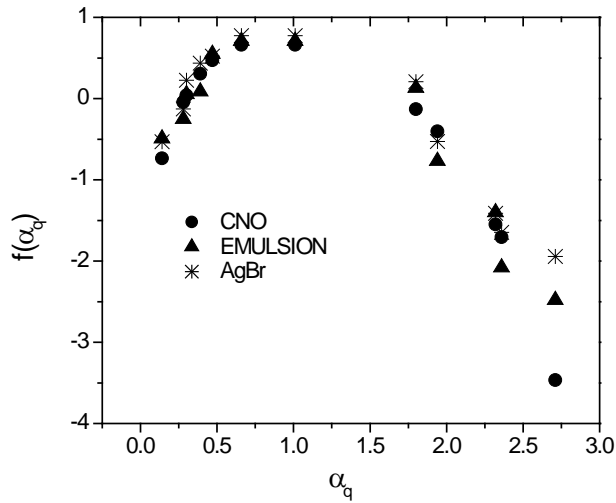

**Figure 5.** Variations of $f(\alpha_q)$ with $\alpha_q$ in 14.5A GeV/c $^{28}$Si-nucleus collisions for various targets.

**Table 1.** Values of Renyi dimensions, $D_q$, in 14.5 A GeV/c $^{28}$Si-nucleus collisions for $q = 0, 1$ and 2.

| Data type | Renyi dimensions | | |
|---|---|---|---|
| | $D_0$ | $D_1$ | $D_2$ |
| Experimental | $0.710 \pm 0.011$ | $0.680 \pm 0.17$ | $0.549 \pm 0.071$ |
| **FRITIOF** | $0.790 \pm 0.011$ | $0.660 \pm 0.005$ | $0.535 \pm 0.032$ |